\documentclass[aps,prd,12pt,preprint,tightenlines,unsortedaddress,
amsfonts,amssymb,amsmath,byrevtex,showpacs,nofootinbib]{revtex4-1}

\usepackage{latexsym}
\usepackage{graphicx}
\usepackage{geometry}
\usepackage{epstopdf}
\usepackage{color}
\usepackage{caption}
\usepackage{subcaption}

\usepackage{xcolor}
\usepackage{fancyhdr}
\usepackage[applemac]{inputenc}

\usepackage[breaklinks,colorlinks,backref]{hyperref}
\hypersetup{
    colorlinks, 
    linktoc=all, 
    linkcolor=red,  
  citecolor=hyptxt,
  urlcolor=blue}
  \hypersetup{
  citebordercolor=,
  filebordercolor=green,
  linkbordercolor=blue
}
\definecolor{hyptxt}{rgb}{0.7, 0.4, 0.9}

\newcommand{\ud}{\mathrm{d}}
\newcommand{\be}{\begin{equation}}
\newcommand{\ee}{\end{equation}}

\newcommand{\R}{\mathbb R}

\newcommand{\ket}[1]{|\kern.3ex#1\kern.3ex\rangle}
\newcommand{\bra}[1]{\langle\kern.3ex #1 \kern.3ex|}
\newcommand{\scalar}[2]{\langle\kern.3ex #1 \kern.3ex|\kern.3ex#2\kern.3ex\rangle}
\newcommand{\norm}[1]{\|\kern.3ex#1\kern.3ex \|}

\def\lg{\langle }
\def\rg{\rangle }

\def\ud{\mathrm{d}}
\def\mfn{\mathfrak{n}}

\begin{document}

\title{Nonadiabatic bounce and an inflationary phase in the quantum mixmaster universe}

\author{Herv\'{e} Bergeron}
\email{herve.bergeron@u-psud.fr} \affiliation{ISMO, CNRS, Univ. Paris-Sud, Universit\'{e} Paris-Saclay, 91405 Orsay, France}

\author{Ewa Czuchry}
\email{eczuchry@fuw.edu.pl} \affiliation{National Centre for Nuclear Research, Ho{\.z}a 69, 00-681 Warszawa, Poland}

\author{Jean-Pierre Gazeau}
\email{gazeau@apc.univ-paris7.fr}
\affiliation{APC, Univ Paris Diderot, CNRS/IN2P3, CEA/Irfu, Obs de Paris, Sorbonne Paris Cité, 75205 Paris, France}
\affiliation{Centro Brasileiro de Pesquisas Fisicas
22290-180 - Rio de Janeiro, RJ, Brazil }

\author{Przemys{\l}aw Ma{\l}kiewicz}
\email{pmalk@fuw.edu.pl}
\affiliation{National Centre for Nuclear Research,  Ho{\.z}a 69,
00-681
Warszawa, Poland}
\affiliation{APC, Univ Paris Diderot, CNRS/IN2P3, CEA/Irfu, Obs de Paris, Sorbonne Paris Cité, 75205 Paris, France}

\date{\today}


\begin{abstract}
Following our previous paper, Bergeron et al, Smooth quantum dynamics of the mixmaster universe Phys. Rev. D 92, 061302(R)  (2015) \cite{B9A}, concerning the quantization of the vacuum Bianchi IX model and the Born-Huang-Oppenheimer framework, we present a further analysis of the dynamical properties of the model. Consistently with the deep quantum regime, we implement the harmonic approximation of the anisotropy potential. We thus  obtain manageable dynamical equations. We study the quantum anisotropic oscillations during the bouncing phase of the universe. Neglecting the backreaction from transitions between quantum anisotropy states we obtain analytical results. In particular, we identify a parameter which is associated with dynamical properties of the quantum model and describes a sort of phase transition. Once the parameter exceeds its critical value, the Born-Huang-Oppenheimer approximation breaks down. The application of the present result to a simple model of the Universe indicates that the parameter indeed exceeds its critical value and that there takes place a huge production of anisotropy at the bounce. This in turn must lead to a sustained phase of accelerated expansion, an inflationary phase. The quantitative inclusion of backreaction shall be examined in a follow-up paper based on the vibronic approach.
\end{abstract}

\pacs{04.60.Kz, 04.60.Ds, 03.65.-w, 03.65.Sq, 04.20.Dw}

\maketitle

\tableofcontents

\section{Introduction}
It is expected that a generic spacelike singularity of general relativity is oscillatory \cite{BKL,BGIMW}. It is conjectured that at any spatial point  approaching a spacelike singularity, the time derivatives of the gravitational field ultimately dominate the spatial ones and that the dynamics at these points becomes asymptotically well-approximated by oscillatory, spatially homogeneous Bianchi models. It is called the Bielinskii-Khalatnikov-Lifshitz (BKL) conjecture. An interesting way of rephrasing the dynamics close to the singularity is to employ the Hamiltonian formalism to bring the asymptotic general relativistic equations to the form of a Fuchsian system (or, Fuchsian-like one) and claim (or, conjecture) the existence of a unique solution with a fixed asymptotic behaviour \cite{Dam}.  The asymptotic  dynamics has been studied and the oscillatory behavior has been confirmed in numerical investigations \cite{numgar}. It was also found that during the evolution towards the singularity there may form and develop sharp features at some points, called spikes \cite{spikes1, spikes2, spikes3}, at which the dynamics is exceptional. 

``Quantization of the BKL scenario'' remains an elusive task.  One can, however, make some progress in understanding the quantum dynamics of the homogenous models and in particular of those which play a pivotal role in the BKL conjecture. This is the aim of the present paper. We are aware that the quantum theory of Bianchi models may be too limited to provide a solid, trustworthy picture of the generic quantum dynamics of the  inhomogeneous gravitational field close to the singular state. Nevertheless, since the Bianchi models are in general anisotropic they contain much richer physics than usually quantized isotropic ones. As a matter of fact, we focus precisely on the role of anisotropic degrees of freedom in the quantized dynamics and in particular on their effect on the quantum bounce.

Among the homogenous cosmological models, the Bianchi IX model (BIX) exhibits the oscillatory behavior on approach to the singularity and has been often a focal point of the research into the singularity problem\footnote{Bianchi type VIII model is another Bianchi class A model with confining anisotropy potential and undergoing oscillations. Contrary to type IX model, it does not posses the isotropic limit.}. It describes the non-linear dynamics of a gravitational wave whose energy fuels the contraction of the isotropic geometry. Its first quantization was implemented by Misner \cite{cwm} within the canonical framework. It was confined to highly excited anisotropic oscillations and did not resolve the singularity. A lot of later work followed the Misner approach introducing some extensions or modifications (see \cite{montani} for references). Since then there have also appeared a number of new approaches, e.g.: in \cite{CH} a sum-over-histories generalized quantum theory of BIX was proposed; in \cite{damspin}  the quantum dynamics of a supersymmetric Bianchi IX model was studied; in \cite{phase} quantum corrections to the classical Bianchi IX universe were derived along the lines of the asymptotic safety program; etc. Recently, an approach based on loop quantum gravity was developed to deal with the singularity problem, see e.g. \cite{Bojowald:2003xe}. A recent review of quantum cosmology was given in \cite{Boj2015}. For a useful list of references related to the subject, see also \cite{damspin}.

Our recent result shows that the singularity of Mixmaster universe can be resolved via affine coherent state (ACS) quantization. Such a covariant integral quantization \cite{gazmur16} respects the half-plane geometry for the isotropic canonical variables. The quantum geometrical term, a repulsive potential that is issued from our quantization procedure, prevents the universe to reach the singular state. Moreover, the non-singular dynamics can be analyzed in its deep quantum domain of low anisotropy eigenstates \cite{B9A,B9B}. Specifically, we showed that in this regime the adiabatic approximation can be applied. We derived the explicit bouncing dynamics of the Mixmaster universe within the Born-Oppenheimer (BO) approximation as well as its refined version, the Born-Huang approximation.

The present paper is devoted to an investigation of quantum dynamics of the Mixmaster beyond the adiabatic approximation. The suitable framework for this investigation is ultimately provided by the vibronic approach known from molecular physics (see \cite{B9B}, \cite{yakorny12}). In this approach the anisotropic oscillations are allowed to transit between eigenstates and the effect of transitions on the dynamics of isotropic variables is taken into account. Because such a framework is computationally heavy, we need first a more intuitive understanding of the dynamics, which is achieved with the present paper focusing solely on the quantum anisotropy transitions and neglecting the backreaction. It is also  useful to acquire such an understanding in order to efficiently make numerical computations within the vibronic approach \cite{vibro}.

It is impossible to analyze the anisotropic oscillations in their exact form. In his paper, Misner applied the so called steep wall approximation to the anisotropic potential to simplify the dynamics of the gravitational wave. Being more interested in the deep quantum domain of the model we make use of the harmonic approximation to the same potential. Both approximations describe rather well two extreme dynamical regimes while the intermediate regime is not well-captured by neither of them. Although the strictly quantitative predictions from our model will certainly break down in the other regime, we expect the essential qualitative features of the obtained dynamics to be universal.

We find that the anisotropic degrees of freedom present in the quantized Mixmaster model may be largely excited when the quantum isotropic geometry undergoes a bounce from contraction to expansion. This breakdown of the adiabatic approximation happens for bounces occurring at very small volumes. More specifically, we identify two factors leading to the nonadiabatic bounces. The first one is the initial quantum state of anisotropy set in the contracting branch and characterized  by a large quantum number. Then the large energy of anisotropy fuels the contraction, counteracts the quantum repulsive potential and postpones the bounce until the universe reaches small volumes. The other one is a free parameter of our quantum model, which originates from our quantization procedure. Its value, which determines the strength of the repulsive potential, can be fixed only in confrontation with some observational data. The combined effect of those two factors is quantified in terms of the so-called stiffness parameter. If this parameter is smaller than its critical value, we recover the result agreeing with the Born-Huang-Oppenheimer approximation of our previous paper \cite{B9B}. Otherwise, occurs a completely new phenomenon which is a pure quantum geometry-induced inflationary phase following a quantum bounce. This quantum inflationary phase is initiated and sustained solely by the creation of anisotropy quanta.

Our results include a detailed computation of the scattering matrix for the mixmaster universe bouncing against the quantum repulsive potential. Thus, we provide a framework for investigating any initial or final quantum anisotropy state in the universe. The simplest case comprises contracting universes in a fixed anisotropy eigenstate. Furthermore, we consider a specific example with parameters that are likely to correspond more or less to the physical Universe which is assumed to be well enough described by a radiation filled Bianchi IX model.

Finally, our theoretical framework has some natural mathematical properties. Among other things, they include a free parameter to be fixed in accordance with observational data, time-reparametrization invariance of the semiclassical-quantum dynamical equations and the use of time-dependent quantum oscillator so much employed in the study of linear perturbation theory on curved backgrounds. All those elements are explained in the main body of the paper.

The paper is organized as follows. We begin by recalling the main equation of our previous work on the adiabatic approximation in Sec \ref{es}. Next, in Sec \ref{qu}, we define a refined, nonadiabatic and unitary dynamics of the model by an extended treatment of anisotropic oscillations. We solve the quantum dynamical equation and discuss the most basic observable, i.e. the amplification of the anisotropy amplitude, in Sec \ref{so}. More details on the quantum anisotropy evolution are provided in Sec \ref{tr} in terms of the scattering matrix. We apply our result to speculate about the physical Universe and conclude about the possibility of the inflationary phase in bouncing universe in Sec \ref{univ}. Finally, we conclude in Sec \ref{co}. In the Appendix \ref{valueK} we analyse the strength of the repulsive potential in our model and in the Appendix \ref{appB} we provide a list of symbols used in our paper.

\section{Adiabatic approximation}\label{es}

The complete definition of the Bianchi IX geometry may be found in many papers, e.g. in \cite{cwm} or \cite{B9A}. We recall that the Bianchi IX phase space is six-dimensional. The Misner (real) variables $\beta_{\pm}$ and $p_{\pm}$ describe the motion of anisotropic distortion to the spherical shape of the universe, while (positive) $q=e^{3\Omega/2}$ and (real) $p=\frac{2}{3}e^{-3\Omega/2}p_{\Omega}$ describe its isotropic expansion and contraction. Since the motion of the isotropic part of the geometry is the half-plane $\{(q,p)\, , \, q>0, p\in \R\}$, it is natural to invoke the ``$ax+b$'' affine group of the real line and its representation in defining quantum theory. In \cite{BDGM} it was shown that covariant quantization based on affine coherent states (ACS) shields the boundary of the phase space, $q=0$, by means of a centrifugal potential $\sim 1/q^2$ and thus removes the cosmological singularity.

In our previous papers \cite{B9A,B9B} we have derived a Hamiltonian for Bianchi IX universe ruling the isotropic expansion at a semiclassical level derived from our ACS formalism and the anisotropic oscillations at a canonical quantum level. With our approach,  the semiclassical Hamiltonian constraint for the Mixmaster universe reads as follows (the lapse function is put $\mathcal{N}=-24$):
\begin{equation}\label{main}
\check{C}(q,p)= \frac{9}{4}\left(p^2+\frac{\hbar^2 K(\psi; n_{\pm})}{q^2}\right)+Lq^{\frac{2}{3}}-E_A(N)q^{-\frac{2}{3}}-E_Mq^k\approx 0.
\end{equation}
where the positive coefficient $K(\psi; n_{\pm})$ arises from our ACS quantization procedure, based on a choice of the fiducial vector $\psi$ (see the Appendix A of \cite{B9B} and the Appendix \ref{valueK} of the present paper).  The integer numbers $n_{\pm}$ are the quantum numbers of two modes of anisotropic oscillations in the harmonic approximation, which are fixed in the adiabatic approximation, and $N=n_++n_-$. The term `$Lq^{\frac{2}{3}}$' represents the isotropic intrinsic curvature. The energy of the anisotropic oscillations is given by the $q$-dependent expression
\begin{equation}
\label{anen}
E_A(N)q^{-\frac{2}{3}}=2\sqrt{AB}\hbar(N+1)q^{-\frac{2}{3}},
\end{equation}
which, for fixed $q$, is a spectral value of the quantum anisotropy Hamiltonian:
\begin{equation}\label{ah}
\hat{H}_{A}:=A\frac{\hat{p}_{+}^2+\hat{p}_{-}^2}{q^2}+Bq^{\frac{2}{3}}(\hat{\beta}_{-}^2+\hat{\beta}_{+}^2).
\end{equation}
The term $E_Mq^k$ represents the matter field and the value of $k$ depends on the corresponding equation of state. For the moment we restrict our model to the vacuum case. The value of coefficient $K \equiv K(\psi; n_{\pm})$ depends on whether the Born-Oppenheimer or the Born-Huang approximation is applied. The latter, referred to as $K_{BH}$, includes an extra correction, while the former, referred to as $K_{BO}$, is simpler and will be used in the subsequent analysis. The coefficients $A, B, K_{BO}, K_{BH}$ and $L$ were derived in \cite{B9B} and read: 
\begin{align*}
K_{BO}  =&  \left(\frac{\nu^2}{16}+\frac{\nu}{4}\xi_{21}+\frac{3\nu}{8}\xi_{10}+\xi_{20}\right)\simeq \frac{\nu^2}{16}  \\
   K_{BH} = & K_{BO}+\frac{2}{9}\xi_{20}(\xi_{10})^2\left(n_+^2+n_-^2+n_+ + n_-+3\right)\\
    L = & 36\mfn^2\xi_{\frac{2}{3}1}(\xi_{\frac{5}{3}0})^{\frac{1}{3}}(\xi_{\frac{5}{3}2})^{\frac{2}{3}}\simeq 36\mfn^2 \\
    \sqrt{AB}=& 12\sqrt{2}\mfn(\xi_{10})^{\frac{4}{3}}(\xi_{20})^{\frac{5}{3}}(\xi_{\frac{5}{3}1})^{\frac{1}{2}}\simeq 12\sqrt{2}\mfn
\end{align*}
where $\xi_{rs}=\xi_{rs}(\nu)=\dfrac{\mathrm{K}_r(\nu)}{\mathrm{K}_s(\nu)}$ and $\mathrm{K}_r(\nu)$ is a modified Bessel function depending on a free parameter $\nu>0$. The appearance of this  parameter is proper  to our ACS quantization procedure and can be partially determined through the comparison of our model with the observational data. A mild restriction on $\nu$ has been found in \cite{BDGM}. The value of $\mfn\neq 0$ is the structure constant of the algebra of Killing vector fields in the spatial leaf $\mathbb{S}^3$. It is related the fiducial volume of the universe, $\mathcal{V}_0=\frac{16\pi^2}{\mfn^3}$, and and has no physical significance. In \eqref{main} we have suppressed the constant $\frac{2\kappa}{\mathcal{V}_0}$, where $\kappa=8\pi G$. It will be restored in Sec. \ref{univ}. 

\begin{figure}
\includegraphics[scale=0.8]{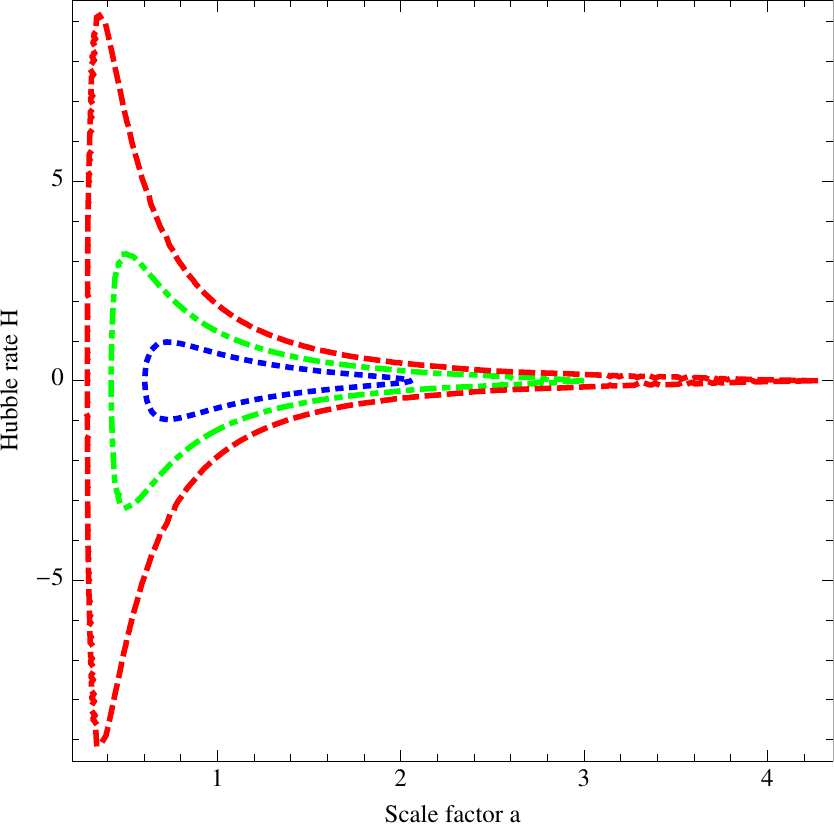}
\caption{Adiabatic dynamics of the BIX universe for $N=0,1,3$, $E_M=0$ and $\nu=2$, where we used: $p=-8Ha^{3/2}$ and $q=a^{3/2}$.}
\label{plot0}\end{figure}

Qualitatively, the dynamics generated by the Hamiltonian constraint \eqref{main} is the following. As the universe contracts and $q$ becomes smaller and smaller, $E_A(N)q^{-\frac{2}{3}}$ grows and the isotropic contraction energy $p^2$ grows with it. Then, at some point, the repulsive term $\frac{\hbar^2 K}{q^2}$ must become dominant as it grows faster than $E_A(N)q^{-\frac{2}{3}}$. Since it comes with the opposite sign, it brings to the halt contraction by forcing $p^2=0$. Then the universe rebounds and the expanding phase begins. See the plot \ref{plot0}.

We note that the eigenenergy $E_A(N)q^{-\frac{2}{3}}$ of anisotropic oscillations is fixed in Eq.\;\eqref{main}. This is true within the adiabatic approximation. The present paper is devoted to a study of conditions under which the adiabatic approximation breaks down and one needs to allow for transitions (excitation or decay) of an initial quantum state of anisotropy. As the anisotropic energy features in the Hamiltonian constraint \eqref{main}, the breakdown of the adiabatic approximation will have an effect on the isotropic background's evolution. However, in what follows we neglect this effect.

\section{Nonadiabatic extension}\label{qu}

\subsection{General setup beyond  BO approximation}

In what follows we  allow for the breakdown of the adiabatic approximation made in \eqref{main}. In other words, we let the energy of anisotropy, $E_A(N)$, vary with time. This can happen in response to sudden and significant changes to the isotropic geometry described by $q$ and $p$. The produced energy of anisotropy would gravitate and influence the evolution of $q$ and $p$. However, we neglect this effect and solve for the dynamics of $q$ and $p$ by keeping $E_A(N_i)$ fixed, where $N_i$ is the initial number of anisotropic quanta. This framework allows to address in a completely analytical manner many interesting questions and thus, understand the mechanism of this phenomenon. For example: (i) What is the regime of validity of the adiabatic approximation? (ii) What are the precise factors on which the excitation (or, decay) of anisotropy depends? (iii) What is the amount of anisotropic energy that can be produced in a violent bouncing cosmological scenario? Etc.

We recall that our semiclassical theory for isotropic mixmaster geometry was derived by first quantizing the constraint and then giving it a semiclassical portrait in the kinematical phase space. As a result, we obtain a semiclassical Hamiltonian constraint theory that is time-reparametrization invariant in the same way as its classical counterpart. We are free to solve the dynamics by means of any time parameter, which is fixed by multiplying the contraint \eqref{main} with a nonvanishing factor, i.e. the lapse function $\mathcal{N}$. Next, Hamilton's equations in a given time parameter are formed by using the Poisson bracket formalism.

Suppose (to be proved later) that the excitation of the eigenstates corresponding to  $n_{\pm}$ occurs only in the vicinity of the bounce, where we assume that the intrinsic curvature term is negligible by putting $L=0$. We transform the Hamiltonian constraint \eqref{main} in order to generate dynamics in the conformal time $\tau$ by changing the choice of lapse to $\mathcal{N}=-q^{\frac{2}{3}}$. We obtain the respective Hamilton equations,
\begin{equation}
\frac{\ud q}{\ud\tau}=\bigg(\frac{q^{\frac{2}{3}}}{24}\bigg)\frac{\partial}{\partial p}\check{C}(q,p),~~~\frac{\ud p}{\ud\tau}=-\bigg(\frac{q^{\frac{2}{3}}}{24}\bigg)\frac{\partial}{\partial q}\check{C}(q,p),~~~\check{C}(q,p)=0,
\end{equation}
the $\tau$-dependence of $q$, 
\begin{equation}
\label{qtau}
q(\tau)=\left(\frac{E_A(N)}{144}\tau^2+\frac{9}{4}\frac{\hbar^2 K}{E_A(N)}\right)^{\frac{3}{4}}\, ,
\end{equation}
after assuming the bounce at $\tau=0$ and $L=0$.

The energy of anisotropy which is put into the semiclassical constraint \eqref{main} is an instantaneous (i.e. for constant $q$) eigenvalue $E_A(N)q^{-\frac{2}{3}}$ of the anisotropic Hamiltonian \eqref{ah}. This is precisely the Born-Oppenheimer approximation which is based on the assumption of a slow variation of $q$. We now go beyond it by allowing the anisotropic wavefunction to respond to the variation of $q$ in conformal time. The new anisotropic wave function is going to include transitions between instantaneous quantum states. The idea of the following calculation is exactly the one that is used in the inflationary scenarios for determining the creation of tensor perturbations. The main difference is that the considered herein gravity waves are homogenous and the background undergoes a nonsingular bounce.

Let us rewrite the classical version of the anisotropic Hamiltonian \eqref{ah} with the new choice for the lapse function, so that it generates the evolution of anisotropy in conformal time:
\begin{equation}\label{HA}
{H}_{A}:=\frac{A}{24}\frac{{p}_{+}^2+{p}_{-}^2}{q^{\frac{4}{3}}}+\frac{B}{24}q^{\frac{4}{3}}({\beta}_{-}^2+{\beta}_{+}^2)
\end{equation} 
Upon rescaling the anisotropic variables $\beta_{\pm}\rightarrow \Delta_{\pm}:=q^{\frac{2}{3}}\beta_{\pm}$, we transform the conjugate momenta as follows $p_{\pm}\rightarrow P_{\pm}:=q^{-\frac{2}{3}}p_{\pm}+\frac{3}{2}\frac{q^{-\frac{1}{3}}p}{A}\Delta_{\pm}$. The corresponding symplectic forms are related as follows:
\begin{eqnarray}
\ud\beta_{\pm}\ud p_{\pm}&=&\ud\Delta_{\pm}\ud P_{\pm}-\ud\tau\ud\left[\left(-\frac{3q^{-\frac{2}{3}}}{32A}\big(p^2+\frac{\hbar^2K}{q^2}\big)+\frac{L}{24}\right)\Delta_{\pm}^2+\frac{q^{-\frac{1}{3}}p}{8}\Delta_{\pm}P_{\pm}\right]\\ \nonumber
&\equiv&\ud\Delta_{\pm}\ud P_{\pm}-\ud\tau\ud {H}_{ext},
\end{eqnarray}
where we assumed that the dynamics of the isotropic variables $q$ and $p$ is independent of the dynamics of the anisotropic variables $\Delta_{\pm}$ and $P_{\pm}$. This is true as in the Hamiltonian constraint (\ref{main}) we have a fixed energy of an initial quantum state of anisotropy. Thus, the anisotropic Hamiltonian in the new variables reads:
\begin{equation}\label{ch}
{H}_{osc}={H}_{A}+{H}_{ext}=\frac{A}{24}(P_{+}^2+P_{-}^2)+\frac{B}{24}\left[1-\frac{9}{4}\frac{\hbar^2K}{AB}q^{-\frac{8}{3}}+\frac{L}{AB}\right](\Delta_-^2+\Delta_+^2)
\end{equation}
As done previously, we neglect the term $L/AB$. Therefore, we describe the anisotropic oscillations in terms of a system of two uncoupled harmonic or anti-harmonic oscillators  both of the same mass and time-dependent frequency:
\begin{equation}
\label{omtau}
m=\frac{12}{A},~~~~\omega^2(\tau)=\frac{AB}{144}\left[1-\frac{9}{4}\frac{\hbar^2K}{AB}q^{-\frac{8}{3}}(\tau)\right].
\end{equation}
We note that there will be times, for small values of $q(\tau)$, when $\omega^2(\tau)<0$ causing amplification or suppression of the amplitude. Our next step is to describe this process at the quantum level. 

\subsection{Nonadiabatic quantum dynamics of anisotropy}

We begin with the classical Hamiltonian given in Eq.\;\eqref{ch}. We proceed with the usual canonical quantization in the Heisenberg picture. This is legitimate since the range of variables $\Delta_{\pm}$ and $P_{\pm}$ is the whole $\R$ and so their respective phase spaces have the Weyl-Heisenberg symmetry. We introduce time-independent creation and annihilation operators
\begin{equation}
\hat{\Delta}_{\pm}=\sqrt{\frac{\hbar}{2m}}\left(v_{\pm}^*(\tau)\hat{a}_{\pm}+v_{\pm}(\tau)\hat{a}_{\pm}^{\dagger}\right)\,. 
\end{equation}
Once they are inserted  into the (quantized) Hamilton equations we obtain
\begin{equation}
\hat{P}_{\pm}=\sqrt{\frac{\hbar m}{2}}\left(\acute{v}_{\pm}^*(\tau)\hat{a}_{\pm}+\acute{v}_{\pm}(\tau)\hat{a}_{\pm}^{\dagger}\right)\,,
\end{equation}
where the acute means, as usual, derivation with respect to the conformal time, and
\begin{equation}\label{master}
\frac{\ud^2 v_{\pm}(\tau)}{\ud\tau^2}=-\omega^2(\tau)v_{\pm}(\tau).
\end{equation}
Upon invoking the canonical commutation rule  $[\hat{\Delta}_{\pm},\hat{P}_{\pm}]=i\hbar\,I$, we obtain that $(v^*_{\pm}\acute{v}_{\pm}-v_{\pm}\acute{v}^*_{\pm})[\hat{a}_{\pm},\hat{a}_{\pm}^{\dagger}]=2i\,I$. Therefore, we impose the normalization condition $v^*_{\pm}(\tau)\acute{v}_{\pm}-v_{\pm}(\tau)\acute{v}^*_{\pm}=2i$ on the solutions to (\ref{master}). 

The quantized Hamiltonian (\ref{ch}) reads $\hat{H}_{osc}= \hat{H}_{osc}^{+}+\hat{H}_{osc}^{-}$ with
\begin{eqnarray}\nonumber
\hat{H}_{osc}^{\pm}= \frac{\hbar}{4}\left(\omega^2(\tau)v_{\pm}^*(\tau)^2+\acute{v}_{\pm}^*(\tau)^2\right)\hat{a}_{\pm}^2+\frac{\hbar}{4}\left(\omega^2(\tau)v_{\pm}(\tau)^2+\acute{v}_{\pm}(\tau)^2\right)\hat{a}_{\pm}^{\dagger~2}\\ \label{qch} +\frac{\hbar}{4}\left(\omega^2(\tau)|v_{\pm}(\tau)|^2+|\acute{v}_{\pm}(\tau)|^2\right)\left(2\hat{a}_{\pm}^{\dagger}\hat{a}_{\pm}+1\right)
\end{eqnarray}
For the $n_{\pm}$-th eigenstate of the number operator, $\hat{n}_{\pm}=\hat{a}_{\pm}^{\dagger}\hat{a}_{\pm}$, we find the expectation values of the Hamiltonians (\ref{qch}):
\begin{eqnarray}\label{eham}
\langle e_{n_{\pm}}|\hat{H}_{osc}^{\pm}|e_{n_{\pm}}\rangle=\frac{\hbar}{2}\left(\omega^2(\tau)|v_{\pm}(\tau)|^2+|\acute{v}_{\pm}(\tau)|^2\right)\left(n_{\pm}+\frac{1}{2}\right)
\end{eqnarray}
This allows to compute the averaged number of particles with respect to the ``instantaneous vacuum"  
\begin{equation}\label{enum}
\langle n_{\pm}\rangle(\tau)=\frac{\left(\omega^2(\tau)|v_{\pm}(\tau)|^2+|\acute{v}_{\pm}(\tau)|^2\right)\left(n_{\pm}+\frac{1}{2}\right)}{2\omega(\tau)}-\frac{1}{2}.
\end{equation}
We set the vacuum state $\hat{a}_{\pm}|0\rangle=0$ to minimize the initial value of Hamiltonian by providing the following initial conditions for $v_{\pm}(\tau)$ (up to a phase):
\begin{equation}\label{vac}
v_{\pm}(\tau_i)=\frac{1}{\sqrt{\omega(\tau_i)}},~~\acute{v}_{\pm}(\tau_i)=i\sqrt{\omega(\tau_i)}.
\end{equation}
In the next section, we construct solutions to the quantum dynamics. 

\section{Determination of the quantum anisotropy evolution}\label{so}
In what follows we solve the previously established quantum equations of anisotropic motion in the semiclassical isotropic background. Then we use the solution to determine the extent of the excitation of anisotropy starting from initial conditions given by a fixed eigenstate, that is, a Born-Oppenheimer-type dynamics in the pre-big bounce era.

\begin{figure}
\includegraphics[scale=0.5]{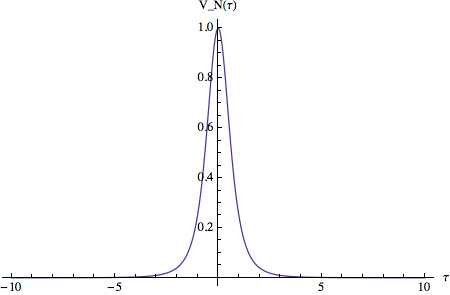}
\caption{Potential $V_N$ for $k_N=1$. It arises and vanishes sharply reflecting a sudden change in the background's dynamics in the vicinity of the bounce.}
\label{1}\end{figure}

The formula (\ref{master}) for $v_{\pm}(\tau)$, supplemented with Eqs.\;\eqref{qtau} and \eqref{omtau}, may be interpreted as the stationary Schr\"odinger equation for a particle with energy $\lambda$ in the potential barrier $V_N$ (see Fig.\;\eqref{1}):
\begin{equation}\label{eom}
\left(-\frac{\ud^2}{\ud\tau^2}+V_N(\tau)\right)v_{\pm}=\lambda v_{\pm},
\end{equation}
with
\begin{equation}\label{V}
\lambda=\frac{AB}{144},~~V_N(\tau)=\frac{1}{\left(k_N\tau^2+k_N^{-1}\right)^2},
\end{equation}
where we $k_N=\dfrac{E_A(N)}{18\hbar\sqrt{K}}$ is the characteristic wave-number and $N=n_-+n_+$ denotes the average number of particles contained in the initial quantum state and featuring in Eq.\;\eqref{main}. The potential $V_N$  arises rapidly, reaches its maximum value $k_N^2$ at $\tau=0$ and vanishes rapidly on timescales $k_N^{-1}$. The solution changes temporarily its character form oscillatory to exponential if $k_N^2>\lambda$, that is, when $N+1>\frac{3}{4}\sqrt{K}$. Therefore, the dynamics of anisotropic variables across the bounce will depend on the bounce's properties, with smaller $K$ and larger $N$ corresponding to a ``stiffer" bounce. The latter reflects the nonlinear nature of the dynamics of anisotropy. For the critical value of $K=\frac{3}{4}$ \footnote{This value was found to provide a unitary dynamics for isotropic models without the need for boundary conditions for the wave function, see \cite{BDGM} for details.}, all quantum states will undergo to some extent the amplification of its amplitude. The important thing is to determine the extent of the excitation due to depth and duration of the negative energy phase of the passing ``particle".

\subsection{Two-regime approximation}
\begin{figure}
\includegraphics[scale=0.5]{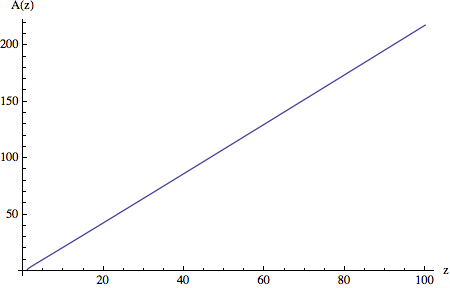}
\caption{Amplification $\mathcal{A}(z)$ obtained within the two-regime approximation. For  $z\gg 1$ we get $\mathcal{A}(z)\approx\pi z$. We note that the amplification is an unbounded function of $z$, and thus, of the initial number of particles $N$.}
\label{3}\end{figure}
Let us assume that $\lambda\ll k_N^2$ (otherwise there is no amplification). Then Eq.\;\eqref{eom} can be split into two regimes: (i) $\lambda\gg V_N$ and (ii) $\lambda\ll V_N$, in which it takes two distinct approximate forms:
\begin{equation}
(i)~-\frac{\ud^2}{\ud\tau^2}v_{\pm}=\lambda v_{\pm},~~~(ii)~\left(-\frac{\ud^2}{\ud\tau^2}+V_N(\tau)\right)v_{\pm}=0.
\end{equation}

The solution to (ii) is determined exactly to be:
\begin{equation}
v_{\pm}(\tau)=\sqrt{k_N\tau^2+k_N^{-1}}\left(a_0+a_1\arctan(k_N\tau)\right)
\end{equation}
where $a_0$ and $a_1$ are constants. The solution to (i) in the left-hand side of the potential reads $v_{\pm}(\tau)=\frac{1}{\lambda^{1/4}}e^{i\sqrt{\lambda}\tau}$ in accordance with conditions (\ref{vac}), whereas it reads $v_{\pm}(\tau)=\frac{C}{\lambda^{1/4}}e^{i\sqrt{\lambda}\tau}+\frac{D}{\lambda^{1/4}}e^{-i\sqrt{\lambda}\tau}$ in the right-hand side of the potential. Inclusion of the appropriate continuity conditions leads to
\begin{eqnarray}\label{C}
C&=&-ie^{-i2z^{-1}\sqrt{z-1}}\left(i+\sqrt{z-1}\right)\left(1+\left(i+\sqrt{z-1}\right)\arctan\sqrt{z-1}\right),\\\label{D}
D&=&i\left(\sqrt{z-1}+z\arctan\sqrt{z-1}\right),
\end{eqnarray}
where we have introduced the parameter
\begin{equation}\label{z}
z=\frac{k_N}{\sqrt{\lambda}}
\end{equation} 
It represents our ``stiffness parameter'', named so for reasons explained below.

Now we can compute the amplification of amplitude $\mathcal{A}(z)$, which is defined as the ratio of the final to initial amplitude $\mathcal{A}(z):=\left|\frac{v_{\pm}(\tau_f)}{v_{\pm}(\tau_i)}\right|$. It reads:
\begin{equation}\label{Az}
\mathcal{A}(z)=\sqrt{2z^2\arctan^2\sqrt{z-1}+4z\sqrt{z-1}+2z-1}\underset{\mbox{at large}\ z}{\sim} \pi z
\end{equation}
where in computing $|v_{\pm}(\tau)|^2=\lambda^{-1/2}\left(|C|^2+|D|^2+2\textrm{Re}\left(C\bar{D}e^{2i\sqrt{\lambda}\tau}\right)\right)$ we time-averaged the oscillatory part. Making use of $\mathcal{A}(z)$ and (\ref{enum}) we obtain:
\begin{equation}\label{N}
\langle N_{\pm}\rangle_{f}+\frac{1}{2}=\left(2z^2\arctan^2\sqrt{z-1}+4z\sqrt{z-1}+2z-1\right)\left(\langle N_{\pm}\rangle_{i}+\frac{1}{2}\right)
\end{equation}
In Fig.\;\eqref{3} we plot the amplification of the amplitude as a function of $z=\frac{k_N}{\sqrt{\lambda}}$. We clearly see that upon lowering the value of $K$ or increasing the value of $N$, the amplification grows unboundedly. In particular, one finds that $\langle N\rangle_f\propto \langle N\rangle_i^3$, that is, the excitation is a non-linear effect.

\subsection{Analytical approximation}
\begin{figure}[t]
\includegraphics[scale=0.8]{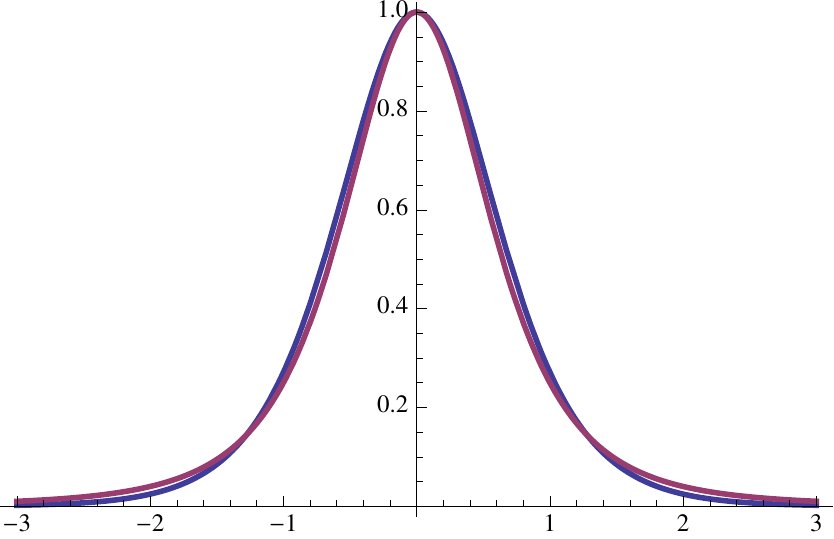}
\caption{The fitting and exact potentials of eq.\;\eqref{V} are hardly distinguishable as shown in the plot.}
\label{fig1}
\end{figure}

In what follows we apply another approximation that is more accurate in describing the amplification of the anisotropy oscillation at low $z$. We will be able to see a phase transition from adiabatic to nonadiabatic behavior. Starting with Eq.\;\eqref{eom} we perform the change of variable $\tau\mapsto u =k_N\tau$ and obtain the new equation:
\begin{equation}\label{neom}
-v_1''(u)+\frac{1}{(u^2+1)^2}v_1(u)=z^{-2}v_1(u)
\end{equation}
where $v_1(u)=v(u/k_N)$. This equation is not explicitly solvable. But we can approximate the potential by a new one which leads to an explicit solution. Namely, we notice that numerically
\begin{equation}
\frac{1}{(u^2+1)^2} \simeq \frac{1}{\cosh^2(4 u /\pi)}
\end{equation}
This is shown in the figure\;\eqref{fig1}.
The coefficient $\alpha  = 4/\pi$ is obtained by imposing
\begin{equation}
\int_{\mathbb{R}} \frac{{\rm d} u}{(u^2+1)^2} = \int_{\mathbb{R}} \frac{{\rm d} u}{\cosh^2(\alpha u)}.
\end{equation}
Therefore we replace Eq. \eqref{neom} by the new one
\begin{equation}
-v_1''(u)+\frac{1}{\cosh^2(4 u /\pi)}v_1(u)=z^{-2}v_1(u) \,.
\end{equation}
Upon introducing another change of variable, $u \mapsto \xi = 4u/ \pi$, we obtain the equation
\begin{equation}
\label{schro3}
-v_2''(\xi) + \frac{( \pi/4)^2}{\cosh^2 \xi} v_2(\xi) = k^2 v_2(\xi) \,,
\end{equation}
where $v_2(\xi)=v_1(\xi\pi/4)$ and $k = z^{-1}\pi/4$. This equation can be solved as follows. With 
\begin{equation}
s := \frac{1}{2} \left(-1+\sqrt{1-( \pi/2)^2} \right) 
\end{equation}
the solution of Eq. \eqref{schro3}, compatible with the expected asymptotic behaviour,  reads as
\begin{equation}
v_2(\xi) = \,  (1-(\tanh \xi)^2)^{ik/2} \, _2F_1\left(ik-s,ik+s+1;1+ik;\frac{1+\tanh \xi}{2} \right)
\end{equation}
up to a multiplicative constant. From the properties of hypergeometric functions, we find the asymptotic behavior
\begin{equation}
\left.
\begin{array}{c}
v_2(\xi) \simeq_{\xi \to -\infty} 2^{ik} \, e^{i k \xi}\\
v_2(\xi) \simeq_{\xi \to +\infty} C \, e^{i k \xi} + D \, e^{-i k \xi}
\end{array}\right.
\end{equation}
with
\begin{equation}
\left.
\begin{array}{c}
C=2^{ik} \, \dfrac{\Gamma(1+ik) \Gamma(ik)}{\Gamma(ik-s) \Gamma(ik+s+1)} \\
D = 2^{ik} \, \dfrac{\Gamma(1+ik) \Gamma(-ik)}{\Gamma(1+s) \Gamma(-s)}
\end{array}\right.
\end{equation}
From Eq. \eqref{enum}, we finally find the excitation $\mathcal{E}(z):=[\mathcal{A}(z)]^2$ as
\begin{equation}\label{excit1}
\mathcal{E}(z) = \frac{N_f+\frac{1}{2}}{N_i+\frac{1}{2}} = 1+2 \frac{|\sin \pi s|^2}{\sinh^2 (\frac{\pi^2}{4z})} = 1+2  \frac{\cosh^2 \pi \chi}{\sinh^2 (\frac{\pi^2}{4z})} \,.
\end{equation}
where $\chi=\Im(s)$. In particular if $z \gg \pi^2/4 \simeq 2.4$ then
\begin{equation}
\mathcal{E} \simeq 1+ \frac{32}{\pi^4} z^2 \cosh^2{\pi \chi}
\end{equation}
We deduce that 
\begin{equation}
\mathcal{E} \gg 1 \quad \text{ if } \quad z \gg \frac{\pi^2}{4\sqrt{2} \cosh(\pi \chi)} \simeq 0.5 \,.
\end{equation}

\noindent The conclusions are the following: (I) there is no maximal value for the excitation $\mathcal{E}$, and excitations occur except for small values of $z$. Recalling the definition of $z$ we find that for
$N_i +1 \gg \frac{3}{8} \sqrt{K}$ excitations occur; (II) there is no excitation when $N_i +1 < \frac{3}{8} \sqrt{K}\,$, this condition being possible only if $K>\frac{64}{9}$, and we recover the adiabatic bounce which was the result of our previous paper \cite{B9B}; (III) since we can prove (see the Appendix \ref{valueK}) that the minimal bound for $K$ is $K \gtrsim 4$, we find that for the minimum value of $K$ excitations occur for all $N_i$, except for the lowest ones, where the excitation is negligible.\\
Assuming $K \simeq  4$, using Eq. \eqref{excit1}, we can express the function $N_f(N_i)$ as
\begin{equation}
N_f = N_i + \cosh^2(\pi \chi) \, \frac{2N_i+1}{\sinh^2 \left(\frac{3\pi^2}{8(N_i+1)} \right)}
\end{equation}
\begin{figure}[t!]
\includegraphics[scale=0.8]{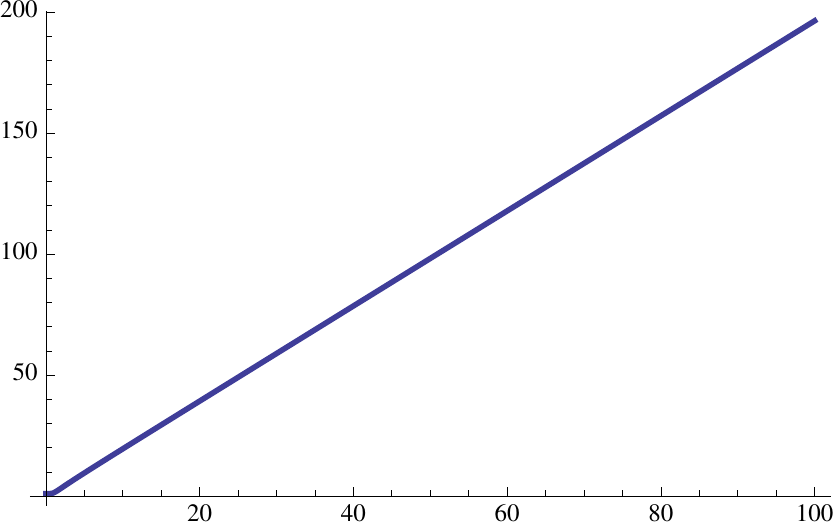}
\includegraphics[scale=0.8]{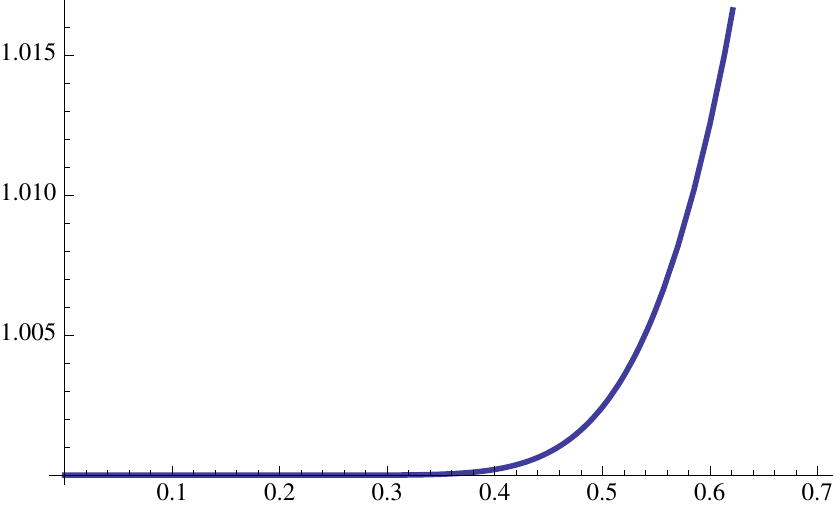}
\caption{Amplification $\mathcal{A}(z)$ of the amplitude of oscillations obtained from the exact solution to the dynamics in the approximate potential. For large values of $z\gg 1$, it coincides with the two-regime approximation (left plot).}
\label{fig2}
\end{figure}
The plot of $\mathcal{A}(z)$ is shown in figures\;\eqref{fig2}. One can see that around $z\approx 0.5$ a kind of phase transition occurs. In order to see it we had to employ the analytical approximation as the phase transition occurs outside the domain of applicability of the two-regime approximation.

\subsection{Numerical investigation and stiffness parameter}

\begin{figure}
\includegraphics[scale=0.45]{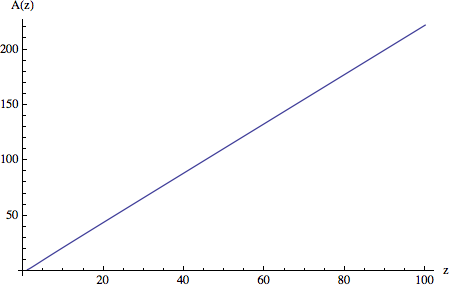}
\includegraphics[scale=0.45]{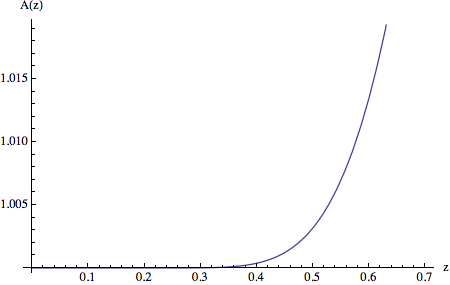}
\caption{Amplification $\mathcal{A}(z)$ obtained from numerical computations. The result coincides with the one obtained from approximating the potential.}
\label{4}\end{figure}

Here we report on  numerical computations that we carried out to describe the structure of the anisotropy excitations caused by the bounce. Let us recall that the excitations  happens because the solution becomes of exponential type during the contraction-expansion transition. It translates into the condition $k_N^2>\lambda$. In Fig.\;\eqref{4} we plot the outcome of numerical computations. We thus confirm the result obtained with the help of our two previous analytical approximation approaches. The result clearly displays the nonlinear nature of the process of excitation of anisotropic distortion oscillations due to the cosmic bounce. The existence of some initial anisotropic distortion will enhance the amplification of  oscillation's amplitude. This process is a purely quantum effect but its nonlinear nature is reminiscent of the classical chaotic behavior of the model on which the quantization was imposed. 

The formulas \eqref{Az} and \eqref{excit1} show that the stiffness parameter $z$ is the only variable in the theory on which the anisotropy excitation depends. Since it is inversely proportional to $K$, the smaller the value of $K$ the more deeply the universe collapses and the more abrupt its rebound is. We see that a bounce with a small enough value of $K$ can turn a smoothly contracting Friedmann-like universe into a highly excited one. Because the final state is expected to be semiclassical, the universe may emerge as highly chaotic into the expanding classical phase. Let us emphasize that $K$ is a purely quantum coefficient, which does not feature in the classical theory. Its very existence is due to our ACS quantization procedure and its value has to be fixed from models consistent with growing observational data. We discuss the possible values of $K$ issued from our quantization in the Appendix \ref{valueK}.

\section{Scattering matrix}\label{tr}
In the previous section, we have dealt with the simplest question concerning the excitation of the energy of anisotropy starting from an eigenstate in the contracting phase. In the present section we develop a framework dedicated to more detailed questions: (i) suppose the universe has started in an anisotropy eigenstate, what is then exactly the final quantum state of anisotropy or (ii) how the dynamics proceeds if the initial state is a generic one? The right choice in this regard is the formalism of the scattering matrix. As we will see soon, the matrix elements depend only on the stiffness parameter $z$.

\subsection{General case}
By introducing the initial condition for $v_{\pm}(\tau)$ in (\ref{vac}) we are in fact defining our creation/annihilation operators as the ones with the corresponding vacuum state minimizing the initial energy. Alternatively, we can set the creation/annihilation operators by requiring that this condition holds for the final state instead of the initial one. Then, in the Heisenberg picture developed here, we obtain a useful Hilbert space basis to discuss a post-bounce quantum state. Suppose that two solutions to (\ref{eom}) are related through a SU(1,1) transform:
\begin{equation}
v_f(\tau)=\alpha v_i(\tau)+\beta \bar{v}_i(\tau)
\end{equation}
where $\alpha$ and $\beta$ are constants such that $|\alpha|^2-|\beta|^2=1$ (to preserve the normalization), and $v_i(\tau)$ and $v_f(\tau)$ denote the solutions with condition (\ref{vac}) fixed before and after the bounce respectively. Then the relation between the respective creation/annihilation operators reads:
\begin{equation}
\hat{a}_i=\bar{\alpha}\hat{a}_f+\beta\hat{a}_f^{\dagger},~~~~\hat{a}_f=\alpha\hat{a}_i-\beta\hat{a}_i^{\dagger}
\end{equation}
where we omit the distinction between the $+$ and $-$ modes. Let us express the initial vacuum state with respect to the final basis as:
\begin{equation}
|0_i\rangle=c_{if}\,e^{\frac{\beta}{2\bar{\alpha}}(\hat{a}_f^{\dagger})^2}|0_f\rangle,
\end{equation}
where $c_{if}$ is the normalization factor. Now, we compute the transition as
\begin{equation}
\langle m_f|n_i \rangle=c_{if}\,\langle m_f|\frac{(\alpha\hat{a}_f^{\dagger}+\bar{\beta}\hat{a}_f)^n}{\sqrt{n!}}e^{\frac{\beta}{2\bar{\alpha}}(\hat{a}_f^{\dagger})^2}|0_f\rangle
\end{equation}
where $m_f$/$n_i$ denotes $m$-th/$n$-th eigenstate in the final/initial basis. We immediately see that $m_f-n_i$ must be even for a non-vanishing element.

Let us introduce the Schr\"odinger-Glauber-Sudarshan coherent states $|\xi\rangle=e^{-\frac{|\xi|^2}{2}}\sum_n\frac{\xi^{n}}{\sqrt{n!}}|n\rangle$. Then,
\begin{eqnarray}\label{cscs}
\langle \xi_f|\xi_i\rangle:=\sum_{m_f,n_i}\frac{\xi_f^{m_f}}{\sqrt{m_f!}}\frac{\xi_i^{n_i}}{\sqrt{n_i!}}\langle m_f|n_i\rangle=
e^{-\frac{|\xi_f|^2+|\xi_i|^2}{2}}N\langle \xi_f|e^{\xi_i(\alpha\hat{a}_f^{\dagger}+\bar{\beta}\hat{a}_f)}e^{\frac{\beta}{2\bar{\alpha}}(\hat{a}_f^{\dagger})^2}|0_f\rangle
\end{eqnarray}
We will place all the creation operators to the left. Making use of the Baker-Campbell-Hausdorf formula we find:
\begin{equation}
e^{\xi_i(\alpha\hat{a}_f^{\dagger}+\bar{\beta}\hat{a}_f)}=e^{\frac{1}{2}\xi_i^2\alpha\bar{\beta}}e^{\xi_i\alpha\hat{a}_f^{\dagger}}e^{\xi_i\bar{\beta}\hat{a}_f}
\end{equation}
We also find:
\begin{eqnarray}
e^{\xi_i\bar{\beta}\hat{a}_f}e^{\frac{\beta}{2\bar{\alpha}}(\hat{a}_f^{\dagger})^2}e^{-\xi_i\bar{\beta}\hat{a}_f}=e^{\frac{\beta}{2\bar{\alpha}}(\hat{a}_f^{\dagger}+\xi_i\bar{\beta})^2}\end{eqnarray}
Inserting the above into (\ref{cscs}) we obtain
\begin{equation}
\langle \xi_f|\xi_i\rangle=e^{-\frac{|\xi_f|^2+|\xi_i|^2}{2}}Ne^{\frac{1}{2}\xi_i^2\alpha\bar{\beta}}e^{\xi_i\alpha\bar{\xi}_f}e^{\frac{\beta}{2\bar{\alpha}}(\bar{\xi}_f+\xi_i\bar{\beta})^2}
\end{equation}
where we used the well-known property: $\hat{a}|\xi\rangle=\xi|\xi\rangle$ (and $\langle\xi|\hat{a}^{\dagger}=\langle\xi|\bar{\xi}$ ). Up to a phase, we determine
\begin{equation}\nonumber
N=\frac{1}{\sqrt{\sum\frac{(2n)!}{(n!)^2}\left(\frac{1-|\alpha|^{-2}}{4}\right)^{n}}}=\frac{1}{\sqrt{|\alpha|}}
\end{equation}
We finally obtain:
\begin{equation}\label{xixi}
\langle \xi_f|\xi_i\rangle=\frac{e^{-\frac{|\xi_f|^2+|\xi_i|^2}{2}}}{\sqrt{|\alpha|}}\exp\left[\xi_i^2\left(1+\left|\frac{\beta}{\alpha}\right|^2\right)\frac{\alpha\bar{\beta}}{2}+2\xi_i\bar{\xi}_f\left(1+\left|\frac{\beta}{\alpha}\right|^2\right)\frac{\alpha}{2}+\bar{\xi}_f^2\frac{\beta}{2\bar{\alpha}}\right]
\end{equation}
In what follows the identification $\alpha=C$ and $\beta=D$ according to Eqs\;\eqref{C} and (\ref{D}) is made. We note that the matrix defined by \eqref{xixi}  is not  unique. We have instead  a continuous family of matrices parametrized by the stiffness parameter $z$ (as we have $\alpha(z)$ and $\beta(z)$). To compute the scattering amplitude correctly one needs to determine the average values of $n_+$ and $n_-$ for the initial state and set the value of $z$. Therefore the scattering against the bounce is not a linear process but a non-linear one in this framework. The superposition principle is broken due to the split between semiclassical isotropic evolution and quantum anisotropic one. Therefore, the dynamics is not, in general, invertible. To obtain a unitary invertible dynamics of all the degrees of freedom one needs to include the backreaction on the background geometry, which is the subject of our future papers devoted to the vibronic approach \cite{vibro}.

\subsection{Case $z\gg 1$}
Let us now inspect formula (\ref{xixi}) under the condition $z\gg 1$. From careful examination of the expansions of $\frac{\beta}{\bar{\alpha}}$, $\left|\frac{\beta}{\alpha}\right|$, $\alpha\bar{\beta}$ and $\alpha$ we find the asymptotic behaviour:
\begin{align}\nonumber
|\langle \xi_f|\xi_i\rangle|\simeq \sqrt{\frac{2}{\pi z}}\exp \left[-\frac{\pi^2z^2}{4}\left((\xi_i^R)^2-(\xi_i^I)^2\right)-\pi z\xi_i^R\xi_i^I\right]\\
\times\exp \left[-(\xi_f^I)^2-\frac{2}{\pi^2z^2}(\xi_f^R)^2-\frac{1}{\pi z}\xi_f^R\xi_f^I\right]\\ \nonumber
\times\exp \left[-\left(\xi_i^R\xi_f^R+\xi_i^I\xi_f^I\right)-\frac{\pi}{2}\left(\xi_i^R\xi_f^I-\xi_i^I\xi_f^R\right)\right]
\end{align}
where $\xi^R=\mathrm{Re}(\xi)$ and $\xi^I=\mathrm{Im}(\xi)$.

\subsection{Excitation from vacuum}
Let us put $\xi_i=0$, that is, the quantized anisotropy is in its vacuum state initially. Then, after the bounce,
\begin{equation}\label{vacexc}
|\langle \xi_f|0_i\rangle|\simeq \sqrt{\frac{2}{\pi z}}\exp \left[-(\xi_f^I)^2-\frac{2}{\pi^2z^2}(\xi_f^R)^2-\frac{1}{\pi z}\xi_f^R\xi_f^I\right]
\end{equation}
The occupation of highly-excited states decreases exponentially with $|\xi_f|^2$. We notice the possibility of the following cosmological scenario: The quantum universe undergoes a smooth adiabatic Friedmann-like contraction phase with anisotropy remaining in a fixed low eigenstate. A surge of the quantum repulsive potential force causes a sudden bounce, which excites the anisotropic oscillations to a large degree. It is almost opposite to what happens according to the classical dynamics, where the growth of anisotropy during contraction is so large that eventually leads to a chaotic contraction. For this reason, cosmological bouncing scenarios aimed at explaining a smooth initial state of the expanding universe tend to employ exotic forms of matter in the contracting phase \cite{NT, PP} to suppress the chaos. Nevertheless, as we show, even in this case a stiff enough bounce will excite anisotropies (a quantum effect) so that the expanding universe may emerge in a chaotic state.

\section{Inflationary bouncing Universe}\label{univ}
In what follows we employ the obtained results to analyze a simple model of the physical Universe. We assume that the whole Universe is well-modelled by a single Bianchi IX model. In our rough estimate we assume that the volume of the present Universe is $V_0=10^{78}\gamma\,\mathrm{m}^3$, where $1<\gamma<\infty$ \footnote{$\gamma=1$ corresponds to the size of observable Universe. However, in order to comply with the requirement of the intrinsic curvature being negligible we need $\gamma\gg1$.}, and the Hubble rate is $H_0\simeq 10^{-26}\,\mathrm{m}^{-1}$. The radiation-matter equality era happened at the redshift $z_e=3500$ (see e.g.\cite{LL}) and thus, $V_e\simeq 10^{67}\gamma\, \mathrm{m}^{3}$ and $H_e\simeq 10^{-24}\,\mathrm{m}^{-1}$. 

\subsection{Adiabatic bounce}
Let us rewrite the semiclassical constraint \eqref{main} in terms of the Hubble rate $H$ and the volume $V=\mathcal{V}_0a^3$. This can be done by using $p=-4Ha^{3/2}\dfrac{\mathcal{V}_0}{\kappa}$, $q=a^{3/2}$ and restoring the constant $\kappa$:
\begin{equation}\label{main2}
H^2+\frac{\kappa^2}{16}\frac{\hbar^2K}{\mathcal{V}_0^2a^6}+\frac{1}{144}\frac{L}{a^2}-\frac{1}{144}\frac{E_A(N)}{a^4}-\frac{1}{144}\frac{E_M}{a^{3/2k-3}}=0
\end{equation}
To consider the earliest phase of the expanding Universe we neglect the isotropic curvature $L=0$ and assume that the matter content is radiation, $k=-2/3$. Then we compute that the number of e-folds $N_{\mathrm{e-f}}$ of the inflationary phase that starts at the bounce for $a_b:~\dot{a}(a_b)=0$ and lasts until $a_f:~\ddot{a}(a_f)=0$. It is in fact independent of the coefficients featuring in \eqref{main2}:
\begin{equation}
N_{\mathrm{e-f}}=\ln\left(\frac{a_f}{a_b}\right)=\ln\sqrt{2}
\end{equation}
According to the standard inflationary scenario, the inflation lasts for at least $N_{\mathrm{e-f}}\simeq 60$ e-folds. We conclude that within the adiabatic approximation (FRW cosmology) there is practically no inflation.

\subsection{Nonadiabatic bounce}
As it was mentioned above,  we put $k=-2/3$ in \eqref{main2} as we focus on the radiation domination epoch that in our scenario lasts until the radiation-matter equality era. At that time the Universe is well in its classical phase and the repulsive potential has become negligible. We also keep the isotropic curvature negligible throughout the epoch, $L=0$.

Plugging the values for the radiation-matter equality era into Eq.\;\eqref{main2} we find that
\begin{equation}
E_A(N)+E_M=\frac{144H_e^2V_e^{4/3}}{\mathcal{V}_0^{4/3}}\simeq 10^{40} \left(\frac{\gamma}{\mathcal{V}_0}\right)^{4/3}
\end{equation}
where $E_A(N)\simeq \frac{10^{-64.4}}{\mathcal{V}_0^{4/3}}(N+1)$. Let us assume low $N$ in the contracting branch and thus $E_A(N)/E_M\simeq 0$,
\begin{equation}
E_M\simeq 10^{39.5} \left(\frac{\gamma}{\mathcal{V}_0}\right)^{4/3}
\end{equation}
Now we are able to derive the stiffness parameter \eqref{z} characterizing the subsequent bounce:
\begin{equation}
\label{zgamK}
z=\frac{2E_M}{3\hbar\sqrt{K}\sqrt{AB}}\simeq 10^{105}\frac{\gamma^{4/3}}{\sqrt{K}}
\end{equation}
where we put $E_M$ instead of $E_A(N)$ which appeared in the previous definitions of $z$ neglecting the matter (see the definition of $k_N$ below Eq.\,\eqref{V}).  We find by means of the two-regime approximation given by Eq.\;\eqref{N} that if the Universe is contracting smoothly with the two modes of anisotropies in their vacua, then after the bounce we get the following number of anisotropy quanta:
\begin{equation}
\lg N_f\rg\simeq z^2\simeq 10^{209}\frac{\gamma^{8/3}}{K}
\end{equation}
and hence the produced anisotropy energy
\begin{equation}
E_A(\lg N_f\rg)\simeq \frac{10^{144.6}}{\mathcal{V}_0^{4/3}}\frac{\gamma^{8/3}}{K}\simeq 10^{105}\frac{\gamma^{4/3}}{K}E_M
\end{equation}
It seems reasonable to expect that the value of $K$ will not be much larger than its smallest possible value allowed by the quantization procedure, i.e. of the order of unity (see Appendix \ref{valueK}). On the other hand, $\gamma\gg 1$. We conclude that just after the bounce the amount of anisotropy $E_A(\lg N_f\rg)$ greatly surpasses the amount of radiation $E_M$ and thus takes over the evolution of the background. In the vicinity of the bounce there must take place a huge production of anisotropy. By the virtue of the constraint equation (\ref{main2}), the growth of the energy of anisotropy has to be balanced by the growth of the isotropic expansion energy. In other words, the behavior of the averaged scale factor will be steeper  in the post-bounce phase. This huge transfer of energy to the isotropic expansion may produce a more or less lasting inflationary phase, which could provide a viable model of the super-accelerated phase in the early Universe in an alternative way to the postulated scalar-driven inflation.

\begin{figure}
\includegraphics[scale=0.5]{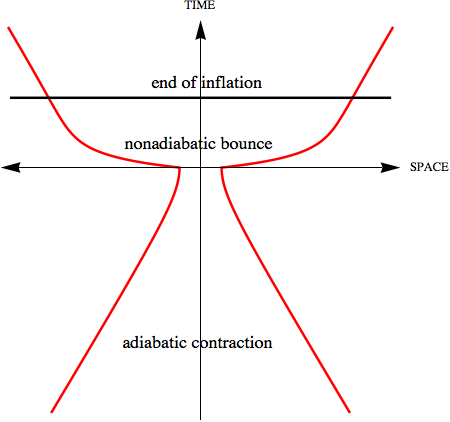}
\caption{The Universe filled with radiation starts in adiabatic contraction with anisotropy in an eigenstate. Then, due to the repulsive potential, the bounce occurs and the anisotropy gets amplified. The produced anisotropy sources an inflationary phase, $\ddot{a}>0$, occurring just after the bounce. Later on, the anisotropy vanishes as $a^{-6}$ and the radiation again dominates the dynamics.}
\label{scenario}\end{figure}

This inflationary phase will last for a finite number of e-folds. To be able to determine this number precisely we must ultimately employ the vibronic approach that is needed to address the post-bounce dynamics quantitatively. However, let us note that once the excitation of the anisotropy is completed the quantum anisotropy state is far from an eigenstate. On the basis of Eq.\;\eqref{vacexc} we infer that the production of the Schr\"odinger coherent states reads:
\begin{equation}
|\langle \xi_f|0_i\rangle|\simeq 10^{-57}\frac{K^{1/4}}{\gamma^{2/3}}\exp \left[-10^{-210}\frac{K}{\gamma^{8/3}}\left(\xi_f^R\right)^2\right].
\end{equation}
Then the Universe must begin to isotropize as the (semi-)classical anisotropy will vanish as $a^{-6}$. We can compute that from the bounce until the radiation-matter equality era in the expanding branch the scale factor must have increased by
\begin{equation}
\frac{a_e}{a_b}\simeq 10^{75}\frac{\gamma}{\sqrt{K}}
\end{equation}
Therefore, the produced anisotropy must have been suppressed relatively to the radiation by the time of the radiation-matter equality era and we have
\begin{equation}
\left(\frac{a_b}{a_e}\right)^2E_A(\lg N_f\rg)\simeq 10^{-46}\gamma^{-2/3}E_M\end{equation}
where the factor $(\frac{a_b}{a_e})^2$ describes the relative diminishing of anisotropy with respect to radiation. This is consistent with our initial assumption about the radiation-matter equality era, which puts anisotropy negligible. The schematic representation of the scenario is given in Fig. \ref{scenario}.

We expect that due to the high expectation value $\lg N_f\rg$, the final anisotropy state will not be accurately described within the harmonic approximation. In order to investigate qualitatively this inflationary phase, and in particular find out the number of e-folds for this phase, the vibronic approach has to be employed \cite{vibro}. Our first computations within the vibronic approach already confirm the proposed scenario.

\section{Conclusions}\label{co}
A non-singular dynamics of the quantum gravitational field is expected to replace the generic cosmological singularity present in the classical theory. The presented analysis of  the quantized Mixmaster dynamics is to our knowledge the most detailed description of this purely quantum phase and its properties. Already the first steps taken herein reveal a rich and unexpected physics of this most fascinating event in the cosmological evolution. However, the amount of theoretical work which is needed in order to obtain a fully satisfactory model of the quantum bounce is enormous. Our forthcoming paper \cite{vibro} extends the present analysis by including backreaction.

Our main finding is a sort of phase transition in the behavior of the anisotropic distortions. Once a critical value describing stiffness of the bounce is reached, the BO approximation breaks down and a highly non-linear excitation of anisotropic eigenstates takes place throughout the bounce. We considered a scenario in which the Universe is isotropically and smoothly contracting in a FRW-like quantum state. The application of our result to this simple model of the Universe shows that there occurs a large production of anisotropy at the bounce, which in turn leads to some sort of a sustained super-expansion phase similar to the one of the standard inflationary models. The most important limitation of our model comes from the fact that the Universe is assumed to be homogenous at the bounce and that the non-perturbative inclusion of inhomogeneities may alter or even prevent the inflationary phase. This limitation, if confirmed, can be  viewed as being analogous to the known constraints on the initial inhomogeneous conditions from which the standard inflationary dynamics fuelled by a massive scalar field could set off \cite{piran}. We note that our notion of inflation is slightly different than the one employed in the conventional inflationary scenarios. It refers to the accelerated expansion of a semiclassical isotropic geometry rather than a purely classical one. This brings justified questions about the qualitative effects this phase could have, e.g.,  on the cosmological perturbations. We expect that the effect could be similar to that of the conventional inflation to the extent to which the semiclassical variables accurately approximate the fully quantum dynamics. This phase will be further studied within the vibronic approach in our next papers.

\acknowledgments
P.M. was supported by the fellowship ``Mobilno\'s\'c Plus" from Ministerstwo Nauki i Szkolnictwa Wy\.zszego.

\appendix

\section{How much stiff is the bounce?}\label{valueK}

In what follows we examine the stiffness of the bounce in terms of the lower bound of the value of $K\equiv K(\psi)$, which depends on the choice of fiducial vector $\psi$ in our quantization scheme. We follow the notation of Appendix A of our previous work \cite{B9B}. We put $K=K_{BO}<K_{BH}$ to find a lower bound on the stiffness of the quantum bounce responsible for the quantum anisotropy excitation. Let us first recall some facts about the affine coherent state quantization from which the value of $K$ is issued.

\subsubsection*{Fiducial vectors in quantization}

The ACS depend on the data of the fiducial vector $\psi \in L^2(\mathbb{R}^+, dx)$. We associate to $\psi$ coefficients $c_\gamma(\psi)$ defined as
$$
c_\gamma(\psi) = \int_0^\infty \frac{{\rm d} x}{x^{\gamma+2}} |\psi(x)|^2 \,.
$$
The ACS quantization is useful only if the $c_\gamma$ are not infinite for a sufficient large domain of $\gamma$, including $\gamma=-1,0,\dots$. This domain is needed to quantize $q^\beta$ for different powers $\beta$, since
$$
A_{q^\beta} = \frac{c_{\beta-1}}{c_{-1}} \hat{q}^\beta\,.
$$
The vector $\psi$ must be normalized, that is,
$$
c_{-2}(\psi) = 1\,.
$$
The fiducial vector $\psi$ is chosen real, to simplify formula of quantized observables (to obtain formula close to the ones arising  from canonical quantization). In order to impose the canonical commutation rule on the basic operators $A_q$ and $A_p$, i.e. $[A_q,A_p]=i \hbar I$, we require
$$
c_0(\psi) = c_{-1}(\psi)\,.
$$
This constraint can be simply obtained by rescaling the fiducial vector as $\psi(x) \mapsto \lambda^{1/2}\psi\left( \lambda x\right)$. Taking into account the previous conditions ($\psi$ real, $c_{-2}=1$, $c_0=c_{-1}$), the constant $K=K_{BO}$ reads
$$
{K}(\psi) = c_{-3}(\psi)^2 \int_0^\infty {\rm d} x \, \psi'(x)^2 (1+x)
$$

\subsubsection*{Different families of fiducial vectors} 
\begin{figure}[t]
\centering
\begin{subfigure}[t]{0.49\textwidth}
\includegraphics[scale=0.85]{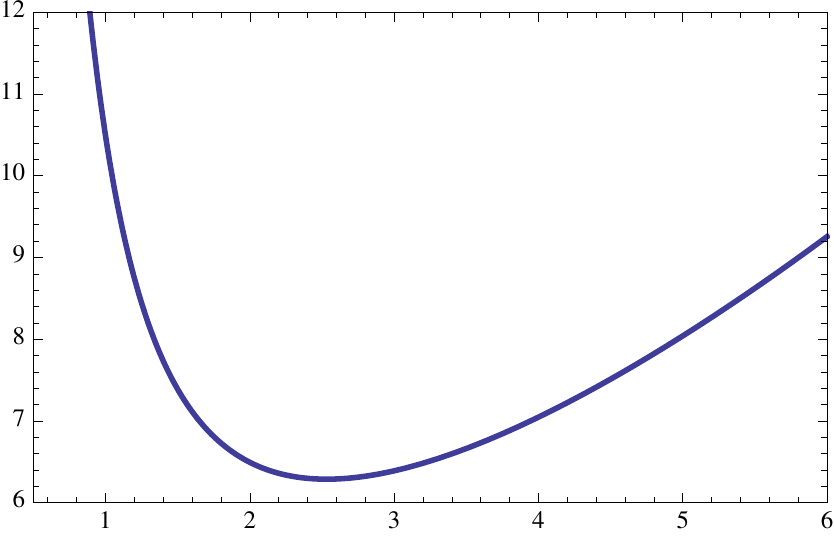}
\caption{${K}(\psi_{\nu})
$ as a function of $\nu$ in accordance with the definition of \cite{B9B}.}
\label{fig3}
\end{subfigure}
\begin{subfigure}[t]{0.49\textwidth}
\includegraphics[scale=0.85]{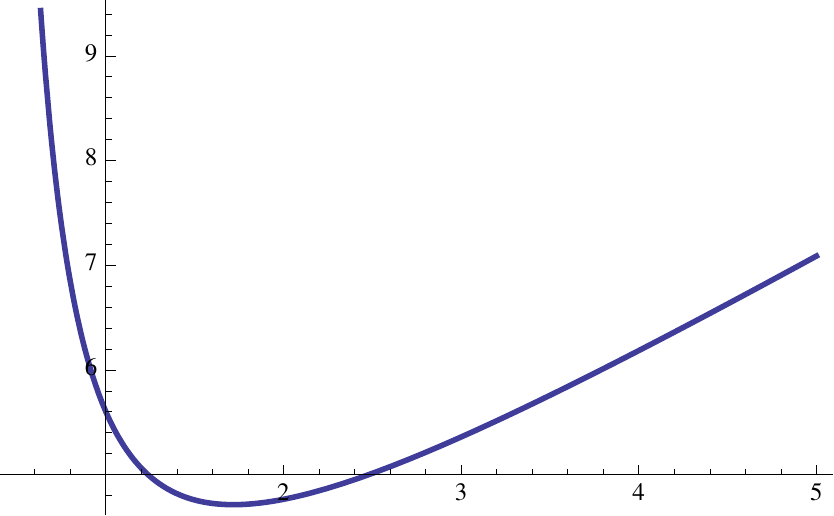}
\caption{${K}(\psi_{\nu})
$  as a function of $\nu$ in accordance with the definition of Eq. \eqref{fctpsi1}.}
\label{fig4}
\end{subfigure}
\caption{The values of parameter K for different $\nu$-dependent families of the fiducial vectors: (a) based on a family introduced in \cite{B9B},~ (b) based on a family introduced in Eq. (\ref{fctpsi1}).}
\end{figure}

We will analyze different choices of the fiducial vector to estimate the lower bound of ${K}(\psi)$. We start with the fiducial vector of our paper \cite{BDGM}  (a function of rapid decrease on $\mathbb{R}^+$). We obtain a function ${K}(\psi_{\nu})
$ plotted in Fig.\;\eqref{fig3}. The minimal value is obtained for $\nu_m \simeq 2.5$, where ${K}(\psi_{\nu_m}) \simeq 6.3$.

We pick another function of rapid decrease, namely
\begin{equation}
\label{fctpsi1}
\psi_{\nu}(x) = \left( \frac{\nu}{\pi} \right)^{1/4} \frac{1}{\sqrt{x}} \exp \left[-\frac{\nu}{2} \left(\ln x - \frac{3}{4 \nu} \right)^2  \right]~.
\end{equation}
The above function, which is nothing but the square root of a Gaussian distribution on the real line with variable $y=\ln x$, centered at $3/4\nu$, and with variance $1/\nu$, verifies the required conditions ($\psi$ real, $c_{-2}=1$, $c_0=c_{-1}$). Furthermore, we have
$$
c_\gamma = \exp \left[\frac{(\gamma+2)(\gamma-1)}{4\nu} \right]~.
$$
In this case, the coefficient ${K}(\psi_{\nu})
$ simply reads
$$
{K}(\psi_{\nu}) = \left(\nu+ \frac{1}{4} \right) \exp \left[ \frac{3}{2\nu} \right]
$$
and is plotted in Fig.\;\eqref{fig4}. The function ${K}(\psi_{\nu})
$ reaches its lower bound at $\nu_m \simeq 1.7$, where ${K}(\psi_{\nu_m}) \simeq 4.7$.

\begin{figure}[t!]
\includegraphics[scale=0.9]{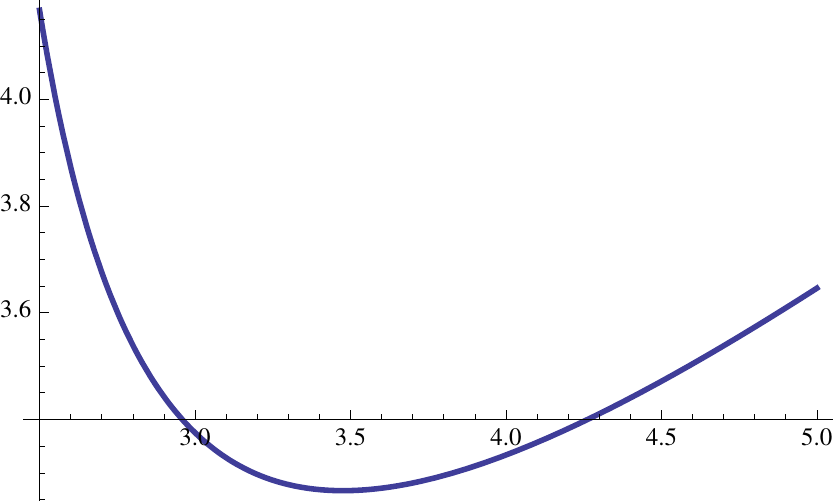}
\caption{${K}(\psi_a)$, issued from $\psi$ of Eq. \eqref{fctpsi2}, versus $\nu$.}
\label{fig5}
\end{figure}

Let us consider a function $\psi(x)$, which is not of rapid decrease at the origin and is vanishing at the origin as a fixed power of $x$, for example
\begin{equation}
\label{fctpsi2}
\psi_a(x) = \frac{(a-1)^{a/2}}{\sqrt{\Gamma(a)}} \, x^{\frac{a-1}{2}} e^{-(a-1) x/2} \quad \text{with} \quad a >2.
\end{equation}
This function verifies the required conditions ($\psi_a$ real, $c_{-2}=1$, $c_0=c_{-1}$). Furthermore we have
$$
c_\gamma = (a-1)^{\gamma+2} \, \frac{\Gamma(a-\gamma-2)}{\Gamma(a)}\,,
$$
where coefficients $c_\gamma$ are well-defined for $\gamma < a-2$. The coefficient ${K}(\psi_a)$ reads
$$
{K}(\psi_a) = \frac{a^2(2a-3)}{4(a-1)(a-2)} \quad \text{for} \quad a>2\,.
$$
The function $K(\psi_a)$ is plotted in Fig.\;\eqref{fig5} and it reaches its lower bound at $a_m \simeq 3.5$, where ${K}(\psi_{a_m}) \simeq 3.3$. 

However, for $a=a_m$ the quantization of $q^\beta$ is possible only for $\beta < a_m-1 \simeq 2.3$. Therefore, this choice of the fiducial vector $\psi$ is much more restrictive than the ones based on a function $\psi$ of rapid decrease. If we require that our procedure of quantization should be defined for all powers of $q$, then the fiducial vector $\psi$ must be a function of rapid decrease and therefore the value ${K}(\psi_{a_m})  \simeq 4$ appears to be a reasonable estimate of the lower bound for ${K}$. We find that at minimal value of ${K}$, the ground state of excitation corresponds to the stiffness parameter $z=\dfrac{2}{3}$, which means that there is practically no excitation and that the adiabatic approximation holds in agreement with the result of our previous paper \cite{B9B}.

\section{Symbols used throughout the text}\label{appB}

\begin{center}
    \begin{tabular}{ | l | p{4cm} | p{8cm} |}
    \hline
    Symbol & Definition & Meaning \\ \hline
    $q,p$ & by \eqref{main}, above \eqref{main2} ~~~~~~~~$p=-4Ha^{3/2}\dfrac{\mathcal{V}_0}{\kappa}$, $q=a^{3/2}$ & isotropic canonical variables \\ \hline
    $K$ & below \eqref{main} & strength of repulsive potential \\ \hline
    $E_A$ & below \eqref{main}~~~~~ $E_A(N)=2\sqrt{AB}\hbar(N+1)$ & energy of anisotropy \\ \hline
    $E_M$ & below \eqref{main} & energy of matter (initially neglected)\\ \hline
    $N,N_i,N_f$ & below \eqref{main} & number of anisotropic quanta (initial, final) \\ \hline
     $A,B,\beta_{\pm},p_{\pm}$ & by \eqref{HA} & constants in anisotropic Hamiltonian, Misner canonical anisotropy variables \\ \hline
    $\Delta_{\pm},P_{\pm}$ & below \eqref{HA} ~~~~~~~~~~~$\Delta_{\pm}=q^{\frac{2}{3}}\beta_{\pm}$, $P_{\pm}=q^{-\frac{2}{3}}p_{\pm}+\frac{3}{2}\frac{q^{-\frac{1}{3}}p}{A}\Delta_{\pm}$ & rescaled anisotropic canonical variables \\ \hline
    $\omega$, $m$ & by \eqref{omtau} $m=\frac{12}{A}$, $\omega^2(\tau)=\frac{AB}{144}\left[1-\frac{9}{4}\frac{\hbar^2K}{AB}q^{-\frac{8}{3}}(\tau)\right]$& frequency and mass of anisotropic oscillations in rescaled variables \\ \hline
    $k_N$ & below \eqref{V} $k_N=\dfrac{E_A(N)}{18\hbar\sqrt{K}}$ & characteristic wavenumber/root of height of scattering potential \\ \hline
    $V_N$ & by \eqref{V} $V_N(\tau)=\frac{1}{\left(k_N\tau^2+k_N^{-1}\right)^2}$ & scattering potential \\ \hline
    $\lambda$ & by \eqref{V} $\lambda=\frac{AB}{144}$ & `energy' of incoming `particle'  \\ \hline
    $z$ & by \eqref{z} $z=\frac{k_N}{\sqrt{\lambda}}$ & stiffness parameter \\ \hline
    $\mathcal{A}(z)$ & below \eqref{z} $\mathcal{A}(z)=\left|\frac{v_{\pm}(\tau_f)}{v_{\pm}(\tau_i)}\right|$ & amplitude amplification \\ \hline
    $\mathcal{E}(z)$ & above \eqref{excit1} $\mathcal{E}(z)=[\mathcal{A}(z)]^2$& anisotropy excitation \\ \hline
    $|\xi\rangle$ & above \eqref{cscs} $|\xi\rangle=e^{-\frac{|\xi|^2}{2}}\sum_n\frac{\xi^{n}}{\sqrt{n!}}|n\rangle$ & Schr\"odinger-Glauber-Sudarshan coherent state \\
    \hline
    \end{tabular}
\end{center}

\end{document}